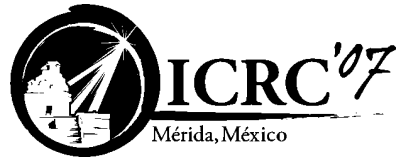

# Highlights from the Pierre Auger Observatory
## – the birth of the Hybrid Era

A A WATSON[1,2]
[1]*School of Physics and Astronomy, University of Leeds, Leeds LS2 9JT, UK*
[2]*Pierre Auger Observatory, Av. San Martín Norte 304, (5613) Malargüe, Argentina*

a.a.watson@leeds.ac.uk

**Abstract:** In this paper some of the highlights from over three years of operation of the Pierre Auger Observatory are described. After discussing the status of the Observatory and over-viewing the potential of the hybrid technique, recent measurements relating to the arrival direction distribution, mass composition and energy spectrum above $10^{18}$ eV are presented. At the time of the presentation at the ICRC no anisotropy had been claimed. From measurements of the variation of the depth of shower maximum with energy, there are indications – if models of high-energy interactions are correct – that the mass composition is not proton-dominated at the highest energies. A flattening of the slope of the energy spectrum from (-3.30 ± 0.06) to (-2.62 ± 0.02) is observed at 4.5 x $10^{18}$ eV while above 3.6 x $10^{19}$eV the flux of cosmic rays is suppressed with the slope becoming (-4.1 ± 0.4). Because of the composition result, caution should be observed over interpretation of the steepening as the long-sought Greisen-Zatsepin-Kuzmin effect. The results are discussed in the context of similar data from the AGASA and HiRes projects and are compared with some models for the propagation of high energy cosmic rays.

## Introduction

Ultra-high energy cosmic rays are of intrinsic interest as their origin and nature are unknown. It is quite unclear where and how particles as energetic as ~ $10^{20}$ eV are accelerated. Over 40 years ago it was pointed out that if the highest energy particles are protons then a fall in the flux above an energy of about 4 x $10^{19}$ eV is expected because of energy losses by the protons as they propagate from distant sources through the CMB radiation. At the highest energies the key process is photo-pion production in which the proton loses about $1/6^{th}$ of its energy in each creation of a $\Delta^+$ resonance. This is the Greisen-Zatsepin-Kuzmin (GZK) effect. It follows that at $10^{20}$ eV any protons observed must have come from within about 50 Mpc and on this distance scale the deflections by intervening magnetic fields in the galaxy and intergalactic space are expected to be so small that point sources should be observed. Despite immense efforts in the period since the prediction, the experimental situation remains unclear.

The main problem in examining whether or not the spectrum steepens is the low rate of events which, above $10^{20}$ eV, is less than 1 per $km^2$ per century so that the particles are only detectable through the giant air showers that they create. These showers have particle footprints on the ground of ~ 20 $km^2$ and suitably distributed detectors can be used to observe them. Also the showers excite molecules of atmospheric nitrogen and the resulting faint fluorescence radiation, which is emitted isotropically, can be detected from distances of several tens of kilometers.

## The status of the Pierre Auger Observatory

The Pierre Auger Observatory has been developed by a team of over 300 scientists and ~100 technicians and students from ~70 institutions in 17 countries. When completed the Observatory will comprise 1600 10 $m^2$ x 1.2 m water-Cherenkov detectors deployed over 3000 $km^2$ on a 1500 m hexagonal grid. This part of the Observatory (the surface detector, SD) is over-looked by 24 fluorescence telescopes in 4 clusters located on four hills around the SD area which is extremely flat. The surface detectors contain 12 tonnes of clear water viewed by 3 x 9" hemispherical photomultipliers. The fluorescence detectors (FD) are designed to record the faint ultra-violet light emitted as the shower traverses the atmosphere. Each telescope images a portion of the sky of 30° in azimuth and 1°- 30° in elevation using a spherical mirror of 3 $m^2$ effective area to focus light on to a camera of 440 x 18 $cm^2$ hexagonal pixels, made of photomultipliers complemented with light collectors, each with a field of view of 1.5° diameter. The status of the Observatory has been described in [1]. For ICRC 2007 data recorded from January 2004 to the end of February 2007 have been analysed. Over this period the number of fluorescence telescopes was increased from 6 to 24 and the number of water tanks from 125 to 1198. Here results from an exposure about 3 times greater than AGASA, and comparable to that of the monocular HiRes detectors at the highest energies, are reported. Above $10^{18}$ eV, more events have been recorded at the Auger Observatory than have come from the sum of all previous efforts. The layout of the instrument is shown in figure 1. As at 9 July 2007, 1438 water-tanks had been deployed, with 1364 currently taking data. All 24 telescopes are working and thus over 85% of the instrument is operational. Except for an area near the centre of the SD array, all landowner issues have been resolved and completion is scheduled for early 2008.

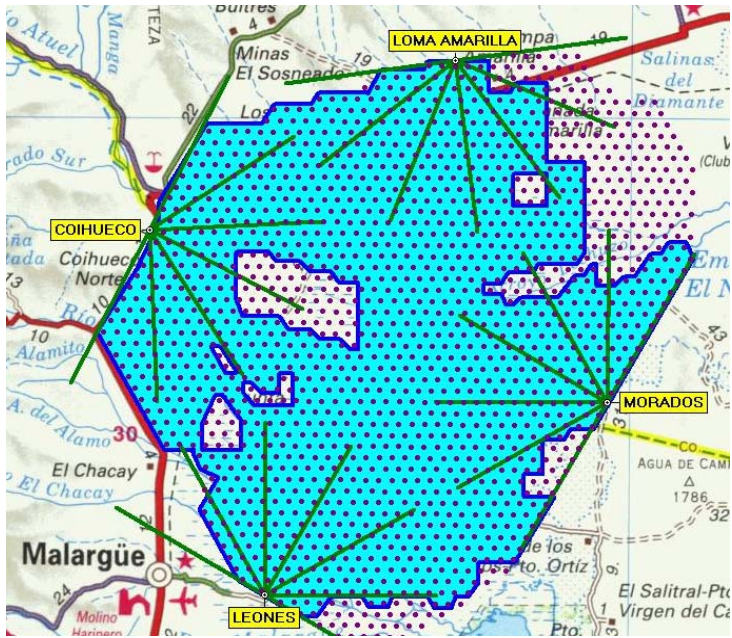

**Figure 1**. The status of the Auger Observatory in early July 2007 is shown. At this time all 24 fluorescence detectors (located at the points marked Leones, Morados, Loma Amarilla and Coihueco) were operational and of 1438 tanks (of a final total of 1600) had been deployed with 1400 filled with water and 1364 taking data. The black dots mark the positions of the water-tanks. See text and [1] for further details.

Undoubtedly one of the highlights of the Observatory is that such a large and multi-national collaboration has succeeded in developing this complex instrument in a rather remote place (Malargüe, Mendoza Province, Argentina) and has used it to produce the results described below in a relatively short time.

An important feature of the design of the Observatory was the introduction of the *hybrid technique* [2, 3] as a new tool to study air-showers. It is used here for the first time. The hybrid technique is the term chosen to describe the method of recording fluorescence data coincident with the timing information from at least one surface detector. The principle is illustrated in figure 2 and the improvement obtained in the determination of $R_p$, the perpendicular distance from the fluorescence detector to the axis of the shower is shown in figure 3. This distance is important when determining the light emitted from the shower axis and when correcting for Rayleigh scattering and for absorption by aerosols. The improvement in the accuracy of angular reconstruction and of the determination of the core position of the shower is about a factor 10 in each case. These conclusions have been obtained empirically using a centrally-positioned YAG laser of 7 mJ at 355 nm.

**Measurements of Arrival Directions**

The search for anisotropies in the arrival directions of cosmic rays has been a goal since their discovery. It has always been expected that as the energy studied increased directional anisotropies would be found although thus far the goal has proved illusive. For the Auger Collaboration early targets have been searches for signals from the galactic centre and for clustering at high energies. Additionally we have looked for effects associated with BL Lacs as have been discussed using northern hemisphere data [4]. None of the earlier claims have been confirmed. Furthermore searches for broad anisotropies, such as dipoles, have served only to set upper limits (e.g. < 0.7% between 1 and 3 EeV), albeit ones that are superior to those of previous studies. More details of the searches made with the Auger database can be found in [5]. When sufficient data have been accumulated it will be possible to make statements about the nature of the arrival direction distribution at the very highest energies.

**Mass composition of the Primary Particles**

The mass composition can be inferred only indirectly by making assumptions about the hadronic interactions at the highest energies. Models of the interactions, such as the SIBYLL or QGSJET families, fit the accelerator data up to the energy of the Tevatron. However extrapolations must be made to the energies of interest here as the centre-of-mass energy in the collision of a $10^{20}$ eV proton with a fixed target is ~30 times that which will be reached at the LHC. Extrapolations of cross-sections, multiplicities, inelasticities etc are necessary. The systematic uncertainties in mass or energy estimates that use hadronic interactions are inherently unknowable but will become better constrained by the LHC, particularly through data from LHCf.

A promising approach is to compare measurements of the depths of shower maxima with the predictions from Monte Carlo calculations using different models of interaction. The maxima can be found to an accuracy of < 20 g cm$^{-2}$ if suitable cuts are made [6]. Each event used in such studies by the Auger Collaboration is a hybrid event in which at least one surface detector has been used to constrain the geometry of the reconstruction. The Auger results are shown in figure 4 where the average of measurements of $X_{max}$ based on 4105 events across two decades of energy are shown together with predictions for proton and iron primaries made using three models of hadronic interactions. It is clear that a single slope does not fit the data and that the rate of increase of $X_{max}$ with energy is smaller above 2 x $10^{18}$ eV than in the region below. The possibility of resolving the question as to whether each data point is associated with a single mass species or with a mixture of masses will be examined by studying the fluctuations in $X_{max}$ at a given energy. Such work is in progress: it requires a more detailed under-standing of systematic uncertainties than is needed for the study of the average behaviour of shower parameters.

In figure 5 the data of figure 4 are compared with those from previous experiments. There is broad agreement although the uncertainties in the earlier work are larger and the energy reach is smaller. For example, in the HiRes report [7], where highest quality stereo events were used, the resolution of $X_{max}$ was 30 g cm$^{-2}$.

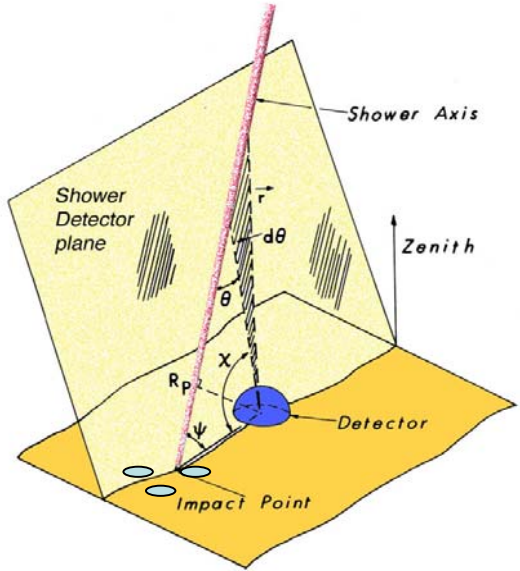

**Figure 2.** The essence of the hybrid technique is illustrated. The shower detector plane is defined by the pattern of pixels that are illuminated at the fluorescence detector. The arrival times of the light lead to an estimate of the orientation of the shower axis in the shower detector plane. This is the procedure adopted for a monocular detector. Using the hybrid method, the time at which the shower axis hits the ground can be deduced from the time at which the shower front hits a surface detector. This time acts as a fitting constraint and allows a greatly improved accuracy of reconstruction over what can be achieved with a monocular system alone or even with a stereo fluorescence pair. The tanks shown close to the impact point have been added to the iconic diagram that is due to the Fly's Eye group (see [8]). There were 419 events above $10^{18}$ eV with the final energy bin centered at 2.5 x $10^{19}$ eV.

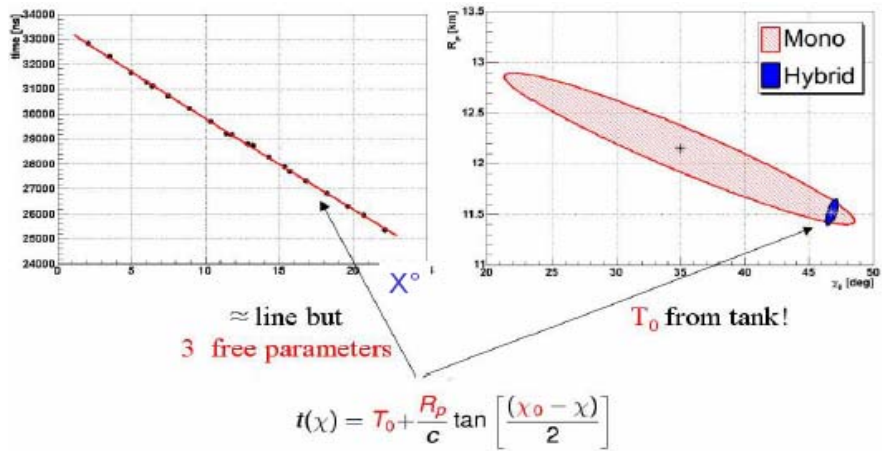

**Figure 3**. The left hand plot shows the times as a function of $\chi$ (see figure 2) from a monocular detector. The solutions for $R_p$ and $\chi_0$ (right hand plot) are degenerate as three parameters must be determined from a line that generally lacks curvature. The hybrid solution (in this case 7 tanks were available) is shown in the right-hand plot: the accuracy is much improved. The diagrams are taken from [9]

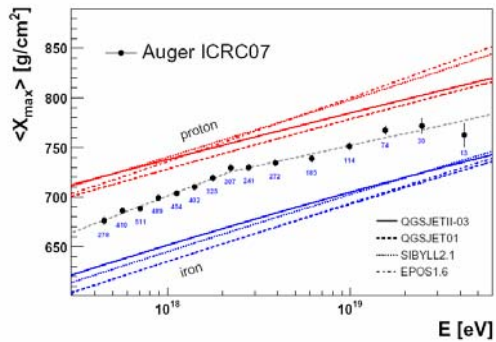

**Figure 4.** The depth of shower maximum, $X_{max}$, as a function of energy. The upper set of lines show predictions made for protons using a range of models: the lower set are made under the assumption of Fe nuclei (taken from [6]).

A preliminary conclusion from the data of figure 4 is that the mass spectrum is not proton-dominated at the highest energies. This unexpected result assumes that the shower models are broadly correct. The position of each data point with respect to the model lines can be used to extract an estimate of $<\ln A>$, where A is the atomic mass (see discussion below relating to figure 10).

Auger data have also been used to set limits to the flux of photons above $10^{19}$ eV. Two methods have been adopted. The first [10] makes use of direct measurements of the depth of shower maximum made with the fluorescence detectors while the second uses the radius of curvature of the shower front and the time-spread of the signals as measured with the surface detectors [11]. The observed distributions are compared with the predictions of Monte Carlo calculations for photon primaries. Showers produced by photons are expected to have larger values of $X_{max}$, curvature and time-spread than showers produced by proton primaries. No photon candidates have been identified and a limit of 2% to the photon flux above $10^{19}$ eV has been set. The significance of this result in the context of 'top-down' models of ultra-high energy cosmic rays is discussed in [11]: most models are ruled out.

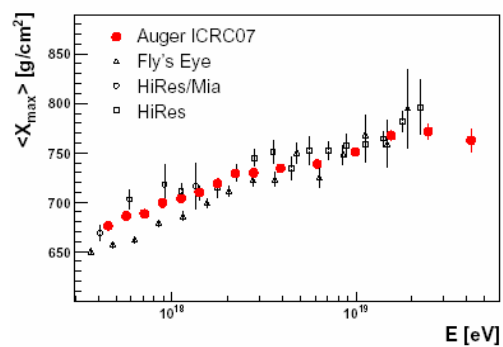

**Figure 5.** Comparison of Auger measurements of the depth of shower maximum with those from previous work (taken from [6]).

A search for tau-neutrinos [12] has been carried out by searching for earth-skimming events at energies above $10^{17}$ eV. Such neutrinos are expected to generate tau-leptons in the earth of sufficient energy to escape and produce showers that could be seen by the surface detectors. So far no tau-neutrinos have been observed and important limits are available in [12].

**The Energy Spectrum**

The hybrid nature of the Auger Observatory enables the energy spectrum of primary cosmic rays to be determined without strong dependence on our limited knowledge of the mass and hadronic interactions. This contrasts with what is required with all-surface detector systems such as were operated successively at Volcano Ranch, Haverah Park, SUGAR, Yakutsk, and AGASA where the use of models was essential for estimates of the primary energy. The Auger approach is to use a selected sample of hybrid events in which the energy can be estimated accurately using the fluorescence detectors. Up to 28 February 2007 there are 357 events that satisfy strict criteria. Direct measurement gives the electromagnetic energy deposited in the atmosphere by the shower which must be augmented by what is carried by high-energy muons and neutrinos which travel into the ground below the atmosphere. This *'missing energy'* must be assessed using assumptions about the primary mass and the hadronic interaction model to give an estimate of the total energy. The correction decreases as the energy increases (largely because of the reduced probability of decay of high-energy pions) and increases as the mass increases. Typical values of the correction [13], assuming a mean mass of 50% protons and 50% iron nuclei, are 19 and 11% at $10^{17}$ and $10^{20}$ eV respectively. At $10^{19}$ eV an average correction is ~12%, with a systematic uncertainty of ± 5%, corresponding to the upper value from EPOS and Fe and the lower value from QGSJET and protons. Thus the derived energy estimates are uncertain by this amount. It is interesting to note that the uncertainty in the energy estimate at $10^{19}$ eV is substantially smaller than in several experiments near the knee region of the cosmic ray spectrum. The most energetic event in the sample has a total energy of 4 x $10^{19}$ eV. For the work discussed here a composition of 50% proton and 50% iron has been adopted along with the QGSJETII model.

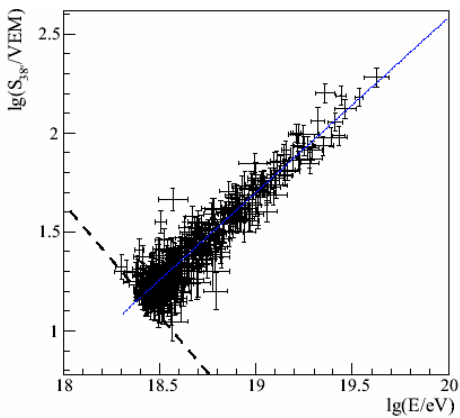

Figure 6. The 387 events with simultaneous measurement of $S_{38°}(1000)$ and an energy derived from the fluorescence detectors (taken from [16]).

The calibration curve, which is used to find the energies of the bulk of the events in which there are only surface detector measurements, is shown in figure 6. The parameter chosen to characterise the size of each SD event is the signal at 1000 m from the shower axes, S(1000), normalised to the mean zenith angle of the events of 38°. The reasons for the choice of S(1000) as the *'ground parameter'* are described in [14]. The method used to combine events of different zenith angle is based on the classical *'Constant Integral Intensity'* method introduced by the MIT group [15]. The tailoring of this approach to the Auger data and the justification for normalisation to 38° are described in [16]. Uncertainties in $S_{38°}$ and $E_{FD}$ are assigned to each event. When determining $E_{FD}$ the absolute fluorescence yield of the 337 nm band in air is taken [17] as 5.05 photons/MeV and the pressure dependence of the fluorescence spectrum is from [18].

For an event to be selected for inclusion in the energy spectrum, the detector in the shower having the highest signal must be enclosed inside an *active* hexagon. An active hexagon is one in which the six surrounding surface detectors were operational at the time of the event. In this way it is guaranteed that the intersection of the axis of the shower with the ground (the shower core) is contained inside the array and therefore that the shower is sufficiently well-sampled to allow a robust measurement of S(1000) and the shower axis. From the analysis of hybrid events, and independently from Monte Carlo simulations, we find that these selection criteria result in a 100% combined trigger and reconstruction efficiency for energies above 3 x $10^{18}$ eV. The area over which the SD events fall and are recorded with 100% efficiency becomes independent of energy above 3 x $10^{18}$ eV. The exposure up to 28 February 2007 is 5165 km² sr yr and is known to 3%. It is about three times that achieved at AGASA and very similar to the monocular HiRes at the highest energies. The energy spectrum derived from nearly 12 000 SD events above $3 \times 10^{18}$ eV is discussed in [16] and is shown here in figure 7.

The slope between $4.5 \times 10^{18}$ and $3.6 \times 10^{19}$ eV is $(-2.62 \pm 0.03)$ based on 5224 events. If this slope is extrapolated to higher energies the numbers expected above $4 \times 10^{19}$ and $10^{20}$ eV are $(132 \pm 9)$ and $(30 \pm 2.5)$, whereas the observed numbers are 51 and 2 respectively. It is thus clear that the slope of the spectrum increases above $\sim 4 \times 10^{19}$ eV, with the significance of the steepening being $\sim 6\,\sigma$. The slope in the highest energy range is $(-4.1 \pm 0.4)$ based on 51 events, 2 of which have energies above $10^{20}$ eV. The limited number of events available for the calibration curve (figure 6) leads to a random uncertainty in the energy scale of 18%. Additionally there is a systematic uncertainty in the fluorescence measurement. This is presently 22% and is dominated by the uncertainty in the fluorescence yield. Measurements are continuing at Frascati and Argonne to improve understanding of the fluorescence yield of this important parameter.

It is clearly desirable to collect more energetic events and this will occur rapidly: the exposure is expected to double within the next 12 months following completion of the Observatory. However an additional exposure of 1510 km$^2$ sr yr (29% of the $\theta < 60°$ data set) is immediately available by using events at large zenith angles ($60 < \theta < 80°$). The size parameter adopted to characterise the horizontal showers is obtained by comparing maps of the observed signal distributions with those from predictions of what is expected at $10^{19}$ eV. The pattern of the maps (see [19] for details) does not depend on the mass or hadronic model and enables the muon number in each event to be estimated. The events are energy-calibrated in the same manner as above using hybrid events but using the muon number as the ground parameter. Presently only 38 inclined hybrid events are available for calibration. The spectrum derived with the inclined showers above $6.3 \times 10^{18}$ eV contains 734 events and the slope above this energy is $(-2.7 \pm 0.1)$.

The spectrum has been extended to lower energies using hybrid data in which at least one tank has registered a signal. Details of this analysis, which takes the spectrum down to $10^{18}$ eV, can be found in [20]: again there is a common calibration for the energy.

The three different spectra are displayed together in figure 8 and checks have shown that the data from these are consistent where they overlap [see 21]. A presentation summarising the situation is shown in figure 9 where the differential intensity J at each energy E is compared with the expectation from a *standard* spectrum. This technique of comparing spectrum data through residuals has been advocated previously [22] and was used in the first presentation of an energy spectrum from the Auger Observatory [23]. The standard spectrum chosen here has a slope of -2.6 and passes through the point at $4.5 \times 10^{18}$ eV which is based on 1631 events.

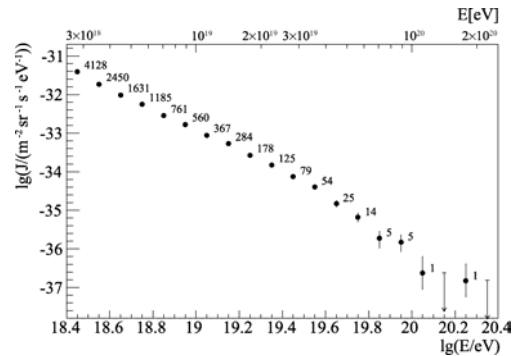

**Figure 7.** The energy spectrum measured with surface detectors using showers with $\theta < 60°$. The spectrum is based on over 12 000 events (taken from [16]).

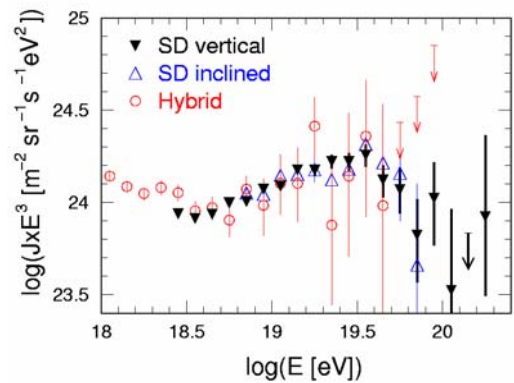

**Figure 8.** Comparison of spectra measured with inclined events ($60 < \theta < 80°$) and with hybrid events. The energy scales are identical so only the statistical uncertainties are shown (taken from [21]).

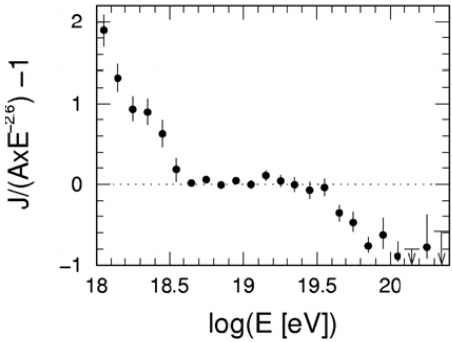

**Figure 9.** The combined Auger spectrum (from fig. 8) shown as residuals from $AE^{-2.6}$ (taken from [21]).

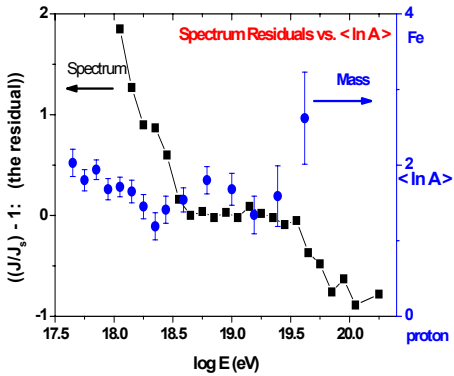

**Figure 10.** The energy spectrum residuals from figure 9 compared with estimates of <ln A> derived by interpolation from figure 4 assuming the QGSJETII-03 model.

An advantage of this style of presentation (figure 9) is that the y-axis is linear. Thus the forgiving nature of a log-log plot is avoided, as are the difficulties of interpreting $J\,E^3$ vs. E plots when the energy scales differ between different measurements. In addition to the steepening of the spectrum at $\sim 4 \times 10^{19}$ eV, an ankle is seen at $4.5 \times 10^{18}$ eV.

In figure 10 the residuals associated with spectrum (figure 9) are shown along with the mean value of ln A (<ln A>) estimated with the QGSJETII-03 hadronic model (see figure 4). With the SIBYLL or EPOS models the estimated values of <ln A> would be larger. Whether there is any significance in the possibility that <ln A> is roughly constant at $\sim 1.6$ in the range where the slope is constant and close to -2.6 is unclear at this stage.

**Comparison with data from AGASA and HiRes**

A preliminary revision of the energy spectrum reported by the AGASA group [24] was presented by Teshima at the RICAP meeting in June 2007. It was stated that there are now 5 or 6 events above $10^{20}$ eV in contrast to the 11 reported previously. The new analysis relies on shower models for a description of the corrections to be made for shower attenuation rather than using the constant integral intensity method. The resulting spectrum is similar to that reported before but features such as the ankle are less evident. As this revision is still preliminary it will not be compared point-by-point with the Auger result but note that the intensities claimed by AGASA are in general significantly higher than those given in [21]. For the differential energy bin centered at $1.12 \times 10^{19}$ eV, the fluxes reported by the two groups differ by $\sim 2.5$.

A point-by-point comparison of the Auger data with that from HiRes I and II, the monocular detectors data [25], is made in figure 11, again using a plot of the residuals. While there appears to be agreement at the highest energies, where there are limited numbers of events, there are differences of up to a factor two between the measurements near $10^{18}$ eV. Such differences cannot be accounted for through observation of different regions of sky and may be associated with the complex aperture calculation required for the HiRes instrument. Assumptions about the primary mass spectrum, the slope of the energy spectrum and the hadronic interaction models are input to this calculation. The aperture is found to change quite rapidly with energy, in contrast to the constant aperture of the Auger Observatory above $3 \times 10^{18}$ eV which is found from knowledge of the lateral distribution and geometry.

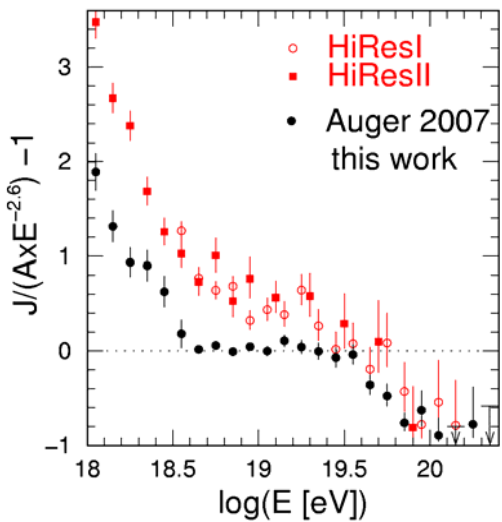

**Figure 11.** A comparison of the residuals of points in the Auger and HiRes spectra. The residuals are with respect to a reference spectrum, $AE^{-2.6}$ [21].

**Comparison of spectra with models of propagation**

Attempts to interpret the shape of the energy spectrum and the evidence about the mass composition involve studying the propagation of an input beam of cosmic rays through intergalactic space. Much attention has recently been given to a model proposed by Berezinsky et al [26] in which it is assumed that the majority of the primaries are protons at the source. These particles lose energy as they propagate through the CMB by photo-pion production and electron pair-production. The features of the spectrum (including what is called a 'dip' in plots of $JE^3$ vs E) reported by the AGASA and HiRes groups are claimed to be well-reproduced by this model [25] although it is made clear in [26] that small mixtures of heavier nuclei (as may be present if the data shown in figures 4 and 10 have been correctly interpreted) would invalidate the argument.

The results of a number of propagation calculations made within the Auger Collaboration are shown in figure 12 (see [21] for details). It is evident that protons models are not good fits to the Auger data in the dip region. Starting with heavy primaries at the source it is apparent that, unlike the case if the primaries were all protons, a second process would need to be introduced to explain the spectrum in the region just above $10^{18}$ eV.

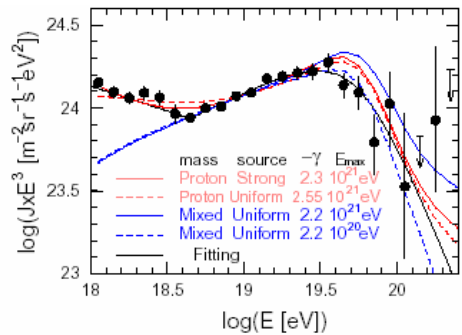

**Figure 12.** The spectra predicted in various propagation models compared with the combined Auger spectrum. The predictions and the models have been normalised at $10^{19}$ eV (taken from [21]).

There is a wide variety of parameters that should be considered in propagation calculations, including the input spectrum slope, the composition at the source, the disposition of the sources, their luminosity and the maximum energy. These are all unknown variables. It is thus not immediately obvious how compelling can be arguments for and against different models that can arise from combinations of many factors. Certainly more data on the mass composition and energy spectrum are needed and discovery of even one point source would have an enormous impact.

**Conclusions from early observations**

The Auger Observatory is over 80% complete and is producing excellent and novel data. The power of the hybrid technique, used for the first time at the Observatory, has been demonstrated. It has led to precise measurements of the depth of maximum as a function of energy and enabled a measurement of the energy spectrum to be made with high statistics and with small reliance on assumptions about hadronic models.

The measurements of the depth of maximum suggest, if the current models of hadronic interactions are correct, that the mass composition is not proton-dominated at the highest energies. An ankle is seen in the energy spectrum at ~ 4.5 x $10^{18}$ eV and a steepening is seen about a decade higher at 3.6 x $10^{19}$ eV. Whether the steepening really is a demonstration of the GZK prediction remains to be seen. The result on primary mass complicates the interpretation and the flux is so low that the anticipated recovery of the spectrum will be hard to observe even with an instrument as large as the Auger Observatory. Two events with energies above $10^{20}$ eV have been detected and the integral flux above $10^{20}$ eV is about 1 per $km^2$ per sr per millennium.

**Plans for the future**

The Pierre Auger Observatory in Argentina is expected to be completed in January 2008 and will be operated for at least ten years. The design for the Observatory, completed in 1995, called for the construction of instruments similar to that shown in figure 1 in both hemispheres. The funding constraints of the late 1990s led to the construction of only the southern arm of the Observatory in Argentina but plans are now well-advanced for the northern section which is targeted for South-east Colorado, USA, near to the town of Lamar. The intention is to construct an array of surface detectors covering 3.5 times the area of the present layout. Details are given in [27]. Submissions to funding agencies are planned for 2008.

It has also been agreed that the energy range studied at the southern section of the Observatory will be increased to collect data down to 2 x $10^{17}$ eV. Studies for two enhancements have been made and funding has been obtained for them. A fluorescence detector system, named HEAT, containing three telescopes, will be constructed near the fluorescence-telescope site at Coiheuco (figure 1) and will cover a range of angles from near 30° to about 60°. The design and planned program of this fluorescence device, which will have three telescopes, is discussed in [28].

On ~ 25 $km^2$ of the Pampa 6 km from, and in the beam of, these additional fluorescence detectors, a complex of water-Cherenkov detectors of the present design, but on a smaller grid-spacing (443 and 750 m), will be deployed. There will also be a set of muon detectors of novel design. These detectors will complement the fluorescence measurements: the water tanks will allow the hybrid technique to be used down to ~ 2 x $10^{17}$ eV and the muon detectors will be used to gain information on the mass composition from this energy up to about 5 x $10^{18}$ eV. Input from the LHC experiments are expected to help resolve the ambiguities arising from the mass/interaction model degeneracy. The instrumentation planned for this extension, named AMIGA, is described in [29] while the science case is set out in [30].

Other methods of studying extensive air-showers are continuously being considered. Within the Collaboration there is an active program to explore again the use of the radio technique to study showers. It has been known since 1965 that radio emission in the 10 – 100 MHz band can be detected and used to explore features of showers. New measurement techniques have been brought to bear on this problem and exploratory work is taking place with various antenna systems at the Auger site. The technique holds the promise of allowing enormous arrays to be instrumented. The current work is described in [31]

**Acknowledgements**

I would like to thank all of my Auger colleagues for their remarkable efforts, but particularly Jim Cronin and Paul Mantsch for without their drive, focus and dedication the Auger enterprise would not have succeeded. I am also extremely grateful to the organisers for inviting me to present the Auger highlights in Mérida.

**References**

[1] B. R. Dawson, for the Pierre Auger Collaboration: 2007, Proc. 30$^{th}$ ICRC (Mérida), #976 [arXiv:0706.1104] and J. Abrahams et al. NIM A 523 50 2004

[2] P. Sommers Astropart. Phys 3 349 1995

[3] B. R. Dawson, H.Y. Dai and P Sommers, Astropart Phys 5 239 1996

[4] C. B. Finley and S. esterhoff for the HiRes Collaboration 2005 Proc. 29$^{th}$ ICRC (Pune):
7 339
Gurbunov D S et al., JETP Letters 80
145 2004

[5] E. Santos, for the Pierre Auger Collaboration: 2007, Proc. 30$^{th}$ ICRC (Mérida), #73, [arXiv:0706.2669]
E.Armengaud, for the Pierre Auger Collaboration: 2007, Proc. 30$^{th}$ ICRC (Mérida), #76, [arXiv: 0706.2640]
D. Harari, for the Pierre Auger Collabor tion: 2007, Proc. 30$^{th}$ ICRC (Mérida), #75, [arXiv:0706.1715]
S. Mollerach, for Pierre Auger Collabora tion: 2007, Proc. 30$^{th}$ ICRC (Mérida), [arXiv: 0706.1749]
J.Abraham et al., Pierre Auger Collaboration, Astropart Phys 27 244 2007

[6] M. Unger, for the Pierre Auger Collaboration: 2007, Proc. 30$^{th}$ ICRC (Mérida), #594 [arXiv:0706.1495v1]

[7] R U Abassi et al., Ap. J. 622 910 2005

[8] J W Keuffel, H E Bergeson and G L Cassiday, Research Proposal to NSF (Fly's Eye), p 17 September 1974
J Bluemer, for the Pierre Auger Collabora tion, Highlight Talk at 28$^{th}$ ICRC (Tsukuba) 2003

[9] M Mostafá, for the Pierre Auger Collabora tion, Nucl. Phys. B (Proc. Suppl.)
165 50 2007

[10] J Abraham et al., Pierre Auger Collaboration, Astropart Phys 27 155 2007

[11] M. Healy, for the Pierre Auger Colla boration, Proc. 30$^{th}$ ICRC (Mérida), # 602 [arXiv: 0710.0025]

D V Semikoz, for the Pierre Auger Collaboration: 2007 Proc. 30$^{th}$ ICRC (Mérida), # 1035, [arXiv:0706.2960]

[12] O Blanch Bigas: for the Pierre Auger Collaboration: 2007 Proc. 30$^{th}$ ICRC (Mérida), # 603, [arXiv:0706.2960]

[13] R. Engel: Rapporteur talk in Proc. 30$^{th}$ ICRC (Mérida), reference to T Pierog et al.

[14] D. Newton, J. Knapp and A. A. Watson, Astropart Phys 26 414 2007

[15] J Hersil et al., Phys Rev Lett 6 22 1961

[16] M. Roth, for the Pierre Auger Collaboration: 2007 Proc. 30$^{th}$ ICRC (Mérida), #313 [arXiv:0706.2096]

[17] M Nagano et al., Astropart Phys 22 235 2004

[18] AIRFLY Collaboration, M Ave et al., Astropart Phys 28 41 2007

[19] P. Facal, for the Pierre Auger Collaboration: 2007 Proc. 30$^{th}$ ICRC (Mérida), #319, [arXiv:0706.4322]

[20] L. Perrone, Pierre Auger Collaboration: 2007 Proc. 30$^{th}$ ICRC (Mérida), #316, [arXiv:0706.2643]

[21] T. Yamamoto, for the Pierre Auger Collaboration: Proc. 30$^{th}$ ICRC (Mérida), paper #318, [arXiv:0706.2638]

[22] A. A. Watson, Nuclear Physics B (Proc Suppl) 22B 116 1991

[23] P. Sommers, for the Pierre Auger Collaboration, Proc.29th ICRC (Pune) 7 387 2005

[24] M. Takeda et al., AstropartPhys 18 135 2003

[25] R.U. Abbasi et al, The HiRes Collaboration: [astro-ph/0703099]

[26] V Berezinsky et al. Phys Rev D 74 040300 2006

[27] D Nitz, for the Pierre Auger Collaboration Proc. 30$^{th}$ ICRC (Mérida), paper # 180, [arXiv: 0706.3940]

[28] H Klages, for the Pierre Auger Collaboration: Proc. 30$^{th}$ ICRC (Mérida), paper #065

[29] A Etchegoyen, for the Pierre Auger Collaboration, Proc. 30$^{th}$ ICRC (Mérida), paper # 1307 [arXiv: 0706.1307]

[30] G Medina Tanco, for the Pierre Auger Collaboration, Proc. 30$^{th}$ ICRC (Mérida), paper #991 [arXiv:0709.0772]

[31] A Van den Berg, for the Pierre Auger Collaboration: Proc. 30$^{th}$ ICRC (Mérida), paper # 0176 [arXiv: 0708.1709